\documentclass[12pt,prd,aps,showpacs,preprint,nofootinbib,floatfix]{revtex4-1}
\pdfoutput=1

\usepackage{epsfig}
\usepackage{graphicx}
\usepackage{multirow} 
\usepackage{color}
\usepackage{hyperref}
\usepackage{amsmath,amssymb}
\usepackage{rotating}
\usepackage[caption=false]{subfig}

\begin{document}

\title{\texorpdfstring{$\alpha_s$}{alpha-s} from the Lattice Hadronic Vacuum Polarisation}

\newcommand\york{Department of Physics and Astronomy, York University, Toronto, Ontario, M3J 1P3, Canada}

\newcommand\yorkmath{Department of Mathematics and Statistics, York University,
Toronto, Ontario M3J 1P3, Canada \\and\\ CSSM, University of Adelaide, Adelaide,
SA 5005 Australia}

\newcommand\aics{RIKEN Center for Computational Science, Kobe, Hyogo 650-0047, Japan}

\author{Renwick~J.~Hudspith}
\email{renwick.james.hudspith@googlemail.com}
\affiliation{\york}
\author{Randy~Lewis}
\email{randy.lewis@yorku.ca}
\affiliation{\york}
\author{Kim~Maltman}
\email{kmaltman@yorku.ca}
\affiliation{\yorkmath}
\author{Eigo~Shintani}
\email{shintani@riken.jp}
\affiliation{\aics}


\date{\today}

\begin{abstract}
We present a determination of the QCD coupling constant, $\alpha_s$, 
obtained by fitting lattice results for the flavour $ud$ hadronic vacuum polarisation function to continuum perturbation theory. We use $n_f=2+1$ flavours of Domain Wall fermions generated by the RBC/UKQCD collaboration and three lattice spacings $a^{-1}=1.79,\, 2.38$ and $3.15\ \text{GeV}$. Several sources of potential systematic error are identified and dealt with.
After fitting and removing expected leading cut-off effects, we find for the five-flavour $\overline{\text{MS}}$ coupling the value $\alpha_s(M_Z)=0.1181(27)^{+8}_{-22}$, where the first error is statistical and the second systematic.
\end{abstract}


\maketitle

\section{Introduction}

The strong coupling constant, $\alpha_s$, of Quantum Chromodynamics (QCD) 
plays a defining r\^ole in the rich dynamics of the theory as it describes 
the interaction strength of the constituent fundamental particles, the 
quarks and gluons. As a key input to the calculation of perturbative 
contributions to physical processes, an accurate determination of
its value is vital for precision tests of the Standard Model (SM).
The lack of sufficient precision has been a bottleneck in accurately 
measuring the Higgs mass and its couplings through the branching ratios 
of $h\rightarrow \bar cc$, $\bar bb$ and $gg$~\cite{Heinemeyer:2013tqa}.
It also represents one of the main uncertainties in determining the cross 
section of the gluon fusion process for Higgs 
production~\cite{Anastasiou:2016cez}. Significant improvements
will, moreover, be required in order to take advantage of the 
few parts per mil determinations of Higgs partial branching fractions
expected from a future ILC~\cite{Peskin:2013xra} to search for evidence
of beyond-the-SM physics in those relative branching 
fractions~\cite{Lepage:2014fla}.

Since $\alpha_s$ runs with scale, it is conventional to facilitate
comparisons by quoting results at a common, agreed-upon scale. We follow 
convention and quote our results for the coupling in the five-flavour 
theory at the Z boson mass scale, $M_Z$, denoted by $\alpha_s(M_Z)$ in 
what follows. The most recent FLAG average~\cite{Aoki:2016frl}
lattice determination is $\alpha_s(M_Z)=0.1182(12)$, and the most 
recent PDG averages~\cite{Olive:2016xmw} $\alpha_s(M_Z)=0.1181(11)$ 
including lattice results and $\alpha_s(M_Z)=0.1174(16)$ excluding them.

It is clear that several lattice determinations dominate the PDG world 
average~\cite{Olive:2016xmw}. Most lattice measurements (including the 
one detailed in this paper) have to match to perturbation theory at an 
available lattice scale where the perturbative expansion is still 
reliable (see, e.g., 
Refs.~\cite{Blossier:2013ioa,Bazavov:2014soa,Chakraborty:2014aca,Maezawa:2016vgv}). 
This can produce complications, since lattice simulations at large scales comparable to their inverse lattice spacing generally have large cut-off effects that must be taken into account. 
On the other hand, these measurements can take advantage of already-generated 
gauge configurations used for other lattice studies. One could also define the coupling non-perturbatively using the Schr\"odinger functional~\cite{Luscher:1993gh,Bruno:2017gxd}
or the Gradient Flow coupling~\cite{Fodor:2012td,Bruno:2017gxd} schemes, which allows access to high matching scales, but these methods
require expensive dedicated simulations. More discussion of recent results 
can be found in the FLAG review~\cite{Aoki:2016frl}. In principle, 
all measurements of $\alpha_s$ should agree within their respective 
systematics. If they do, the combination of complementary yet orthogonal 
determinations should provide the most accurate and dependable evaluation 
of this quantity.

In this paper, we determine $\alpha_s$ by analyzing lattice data for the 
vacuum polarization function (VPF) of the current-current two-point 
function of the flavor $ud$ (isovector), vector current 
$J_\mu =\bar{u}\gamma_\mu d$ using the continuum operator product 
expansion (OPE). This approach is closely related to the most precise of 
the non-lattice determinations, that from finite-energy sum rule (FESR) 
analyses of inclusive hadronic $\tau$ decay data involving exactly the same 
VPF~\cite{Beneke:2008ad,Maltman:2008nf,Boito:2012cr,Boito:2014sta,Pich:2016bdg,Boito:2016oam}.

In the $\tau$ analysis, weighted integrals of the experimental spectral 
distribution from $s=0$ to some upper endpoint $s=s_0\le m_\tau^2$ are 
related by analyticity to integrals of the product of the same weight 
and associated VPF over the circle $\vert s\vert =s_0$ in the complex-$s$ 
plane. For large enough $s_0$, and suitable choices of weight, the OPE 
(possibly supplemented with a large-$N_c$-motivated representation of 
residual duality violations (DVs)) is used to represent the VPF.
The main complication for the $\tau$ analysis comes from the kinematic 
restriction, $s_0\le m_\tau^2$, imposed by the $\tau$ mass. With 
$m_\tau\simeq 1.78$ GeV relatively small, it is not possible to choose 
$s_0$ large enough that higher dimension OPE contributions associated 
with unknown, or poorly known, higher dimension condensates, and residual 
DVs are safely negligible. In fact, if one removes the $\alpha_s$-independent, 
parton-model perturbative contribution from the measured experimental 
spectral distribution, the DV oscillations about the $\alpha_s$-dependent 
perturbative contributions and the perturbative contributions themselves 
are seen to be comparable in size (see, e.g., Fig. 2 of 
Ref.~\cite{Boito:2016oam}). This observation makes clear the necessity of 
attempting to quantify the size of residual DVs in the weighted OPE 
integrals entering the FESR analysis. It is a disadvantage of the $\tau$ 
approach that, at present, this can only be done using models, albeit ones 
constrained by data, for the form of the DV contributions to the spectral 
distribution. 

A key advantage of the alternative lattice approach to analyzing
the same flavor $ud$ VPF is the absence of the kinematic constraint
limiting the $\tau$-data-based analysis. Modern $n_f=2+1$ ensembles 
allow the VPF to be evaluated at (Euclidean) $Q^2$ considerably
larger than $m_\tau^2$. Not only will oscillatory DVs present in the 
spectral distribution at Minkowski $Q^2$ be exponentially suppressed
for Euclidean $Q^2$, but higher dimension OPE contributions can also be
suppressed simply by working at higher $Q^2$ scales. The main 
complications for the lattice approach are (i) understanding at what
$Q^2$ higher dimension OPE contributions become safely negligible, and 
(ii) dealing with and quantifying lattice discretization artifacts 
one expects to encounter at such high scales. Discretization effects can 
(as we will see) be successfully fitted and removed. It is also 
straightforward, in principle, to reduce the errors on $\alpha_s$ 
associated with fitting and removing such discretization artifacts by
suppressing their magnitude through the use, in future, of even 
finer lattices than those employed in the current study. 

The FESR and lattice approaches turn out to be complementary 
when it comes to dealing with complication (i) above. 
The reason is that the freedom 
of weight choice in the FESR framework allows one to perform
multiple analyses with different weights, the different weights producing
integrated OPE expressions with different dependences on the higher
dimension OPE condensates. This freedom can be used to fit the
higher dimension condensates to data. One can then use these results,
now at Euclidean $Q^2$, to determine where such higher dimension OPE 
contributions become safely small, relative to the perturbative 
contributions from which one aims to determine $\alpha_s$. This FESR 
input turns out to be especially important since, with only the VPF
at Euclidean $Q^2$ to work with, it is not possible to reliably
fit higher dimension effective condensates in situations where
multiple such condensates are simultaneously numerically relevant 
and have alternating signs leading to cancellations producing much
weaker $Q^2$ dependence than that of the individual higher dimension
terms. Such a behaviour, which, unfortunately, turns out to be a 
feature of the flavor $ud$ VPF in nature, can lead, at lower
$Q^2$, to combined higher dimension contributions that can mimic
perturbative $Q^2$ dependence and contaminate lower-scale versions 
of the lattice $\alpha_s$ determination. The presence of such a combined
higher dimension contribution is impossible to expose using lattice
data alone. We return to this point in the next section.

In what follows, we will determine $\alpha_s$ using $n_f=2+1$ 
flavour Domain Wall Fermions (DWF) with several lattice spacings. 
This work is built upon a previous determination by one of the authors using different 
ensembles (with Overlap fermions)~\cite{Shintani:2008ga,Shintani:2010ph} where continuum 
OPE results were fit to lattice data with a single lattice spacing
$a$, with $a^{-1} = 1.83$ GeV, in a fit window of $Q$ lying between
$\sim 1.2$ and $1.8$ GeV. We will argue that, because of their
low scale, these earlier analyses were susceptible to the systematic problem 
of contamination by residual higher dimension OPE effects resulting from 
the presence of multiple higher dimension contributions comparable in size 
and with sizable cancellations already discussed above. We stress
that the mimicking of lower dimension contributions by cancelling sums
of multiple higher dimension contributions makes it impossible to rule out
the presence of such contamination using lattice data alone. We believe
this is the reason the result, $\alpha_s(M_Z)=0.1118^{+16}_{-17}$ \cite{Shintani:2010ph},
of this low-scale analysis lies well below the FLAG lattice and PDF 
world averages. We will address these systematics as well as 
several others we have identified, in providing a robust 
determination of $\alpha_s$ using lattice HVP data.

This paper is organized as follows. We first collect some continuum 
perturbation theory results, and outline the FESR analysis, in Sec.~\ref{sec:continuum}. We then briefly describe the 
relevant lattice methodology in Sec.~\ref{sec:lattice}, show the results of fitting 
the continuum perturbative expressions to the lattice data in Sec.~\ref{sec:results}, and conclude the paper in Sec.~\ref{sec:conclusion}.

\section{Continuum results}\label{sec:continuum}

\subsection{Running coupling in Perturbation Theory}

The running coupling, $\alpha_s$, depends on the renormalisation scale $\mu$ and its dependence runs 
with $\mu$ via the renormalisation group equation,
\begin{equation}\label{eq:running}
\mu^2 {\frac{d}{d\mu^2}} \left({\frac{\alpha_s(\mu)}{\pi}}\right) =\, -
\sum_{i=0}\beta_{i}\left( \frac{\alpha_s(\mu)}{\pi}\right)^{2+i}.
\end{equation}
The $\beta$-function is now known at five-loop order in the
$\overline{\text{MS}}$ scheme \cite{Baikov:2016tgj}, but our series coefficients ($A_{ij}$ in App.~\ref{app:aseries}) use only the four-loop $\beta$-coefficients so for  consistency we will use the four-loop running \cite{vanRitbergen:1997va} and three-loop threshold decoupling \cite{Chetyrkin:2005ia} to run our $n_f=3$ results to the five-flavour scale $\mu =M_Z$. In the end, the difference between using the four and five loop running will be small compared to the statistical and systematic errors of our measurement.
The coefficients $\beta_i$ are $n_f$-dependent and for $n_f=3$ are listed in Eq.~\ref{eq:nfthree_beta} of 
Appendix \ref{app:betacoeffs}. The coupling can be run in perturbation theory (PT) from the renormalisation scale $\mu$ to some scale 
$m$ by numerically integrating Eq.~\ref{eq:running}, but 
for scales $m$ close to $\mu$ one can also use the 
series expansion,
\begin{equation}\label{eq:alpha_run}
\frac{\alpha_s(m)}{\pi} = \frac{\alpha_s(\mu)}{\pi} + \sum_{i=2}
\left(\frac{\alpha_s(\mu)}{\pi}\right)^i \sum_{j=1}^{i-1} A_{ij} \ln^j\left(\frac{m^2}{\mu^2}\right).
\end{equation}
The coefficients $A_{ij}$ can be found in Eq.~\ref{eq:ascale} of Appendix \ref{app:aseries}.

\subsection{Adler function and the perturbative series 
for the HVP}

In the continuum, the current-current two-point function, $\Pi_{\mu\nu}$, involving vector or axial vector currents $j^a_\mu(x) \equiv j_\mu (x)$, is defined by
\begin{equation}\label{eq:pidef}
 \Pi_{\mu\nu}(q) = \int d^4x\; e^{iq\cdot x} 
 \left< j_\mu (x) j_\nu^\dagger (0) \right>.
\end{equation}
Generally, this has a Lorentz decomposition into $J=1$ 
(transverse) and $J=0$ (longitudinal) components,
\begin{equation}\label{eq:lorentz}
\Pi_{\mu\nu}(q) = \left( q^2 g_{\mu\nu} - q_\mu q_\nu \right)
\Pi^{(1)}(q^2) - q_\mu q_\nu \Pi^{(0)}(q^2).
\end{equation}
In this work we consider only the flavor $ud$ vector 
current, in the isospin limit. $\Pi_{\mu\nu}$ then 
satisfies an exact Ward identity,
$q^\mu \Pi_{\mu\nu}=0$, and is purely transverse, 
with $\Pi^{(0)}(q^2)=0$. The HVP, $\Pi^{(1)}(q^2)$, will 
be denoted by $ \Pi(q^2)$ in what follows.

The Adler function is a physical quantity related to the 
regularisation- and scheme-dependent HVP by 
$D(Q^2)\equiv -q^2\; \text{d}\Pi (q^2)/\text{d}q^2$,
with $Q^2= -q^2$. $D(Q^2)$ has an OPE representation of 
the form
\begin{equation}
D(Q^2) = D^{(0)}(Q^2,\mu^2) + \frac{m^2}{Q^2}D^{(2)}
(Q^2,\mu^2) +
\sum_{i=2}^\infty {\frac{C_{2i}}{[Q^2]^i}},
\end{equation}
with dimension $D\ge 4$ condensates, $C_D$, parameterising
non-perturbative contributions. At large enough $Q^2$, 
the perturbative (dimension $D=0$) series, $D^{(0)}$,
dominates the HVP and, at five loops, is known to the
highest loop order in continuum perturbation theory. 
The $D^{(2)}$ series, which contains perturbative 
contributions second order in the quark masses,
is known to three loops~\cite{Chetyrkin:1993hi}. In 
Sec.~\ref{sec:FESR}, we will show how to use FESR results 
to help us restrict our attention to $Q^2$ for which 
$D\ge 4$ condensate contributions are safely negligible.
The $D=2$ series is generally strongly suppressed by the 
presence of the squared-light-quark mass factors, but 
does display rather slow convergence. It is, however, 
known, from a comparison of the OPE to lattice data for 
the flavor-breaking $ud-us$ HVP difference, which has
a very similar slowly converging $D=2$ series, that,
despite this slow convergence, the 3-loop-truncated 
version of the $D=2$ series, combined with known 
$D=4$ contributions, provides an excellent representation 
of the lattice data over a wide range of relevant 
Euclidean  $Q^2$~\cite{Hudspith:2017vew}. It is thus possible to 
use the three-loop-truncated form to estimate
$D=2$ contributions to the flavor $ud$ vector HVP, 
and confirm that they are completely negligible for 
all the ensembles considered, at the $Q^2$
we employ in our analysis (see Sec.\ref{sec:FESR}). We will thus be able to analyze
the lattice HVP data using only the the well-behaved,
high-order $D^{(0)}$ part of the continuum OPE series, 
and, moreover, average the results obtained from the
different ensembles. The main task will then be to quantify
residual lattice discretization errors present at these
larger scales.

The five-loop result for the perturbative series
$D^{(0)}$~\cite{Baikov:2008jh,Surguladze:1990tg,Gorishnii:1990vf,Chetyrkin:1979bj},
written in terms of the fixed-scale coupling 
$\alpha_s(\mu )$, is,
\begin{equation}
D^{(0)}(Q^2,\mu^2) = \frac{1}{4\pi^2}\left( d_{00} + \sum_{i=1}
\left(\frac{\alpha_s(\mu)}{\pi}\right)^i \sum_{j=0}^{i-1} d_{ij} t^j\right),
\end{equation}
where $t=\ln(Q^2/\mu^2)$ and the coefficients $d_{ij}$ (for three flavours) are 
given in Eq.~\ref{eq:d0_series} of Appendix \ref{app:aseries}. The series 
for the HVP is then
\begin{equation}\label{eq:hvp_pt}
\Pi(Q^2,\mu^2) = C - \frac{1}{4\pi^2}\left(d_{00} t+\sum_{i=1}
\left(\frac{\alpha_s(\mu)}{\pi}\right)^i \sum_{j=0}^{i-1} d_{ij}
\frac{t^{j+1}}{j+1}\right).
\end{equation}
The unphysical integration constant $C$ carries all of 
the scheme- and regularisation-scale dependence.

It is convenient to introduce the subtracted quantity,
\begin{equation}\label{eq:delta_defn}
\Delta(Q_1^2,Q^2,\mu^2) = -4\pi^2 \left(\frac{
\Pi(Q^2,\mu^2)-\Pi(Q_1^2,\mu^2)}{t-t_1}\right)
 - 1,
\end{equation}
which is independent of the unphysical constant, $C$. The 
normalization is chosen so the leading term in the
perturbative series for $\Delta(Q_1^2,Q^2,\mu^2)$ is 
just $\alpha_s(\mu )/\pi$, precisely the
quantity we wish to measure. The
perturbative series for $\Delta$ has the form,
\begin{equation}\label{eq:delta_series}
\sum_{i=1} \left( \frac{\alpha_s(\mu )}{\pi}\right)^i \sum_{j=0}^{i-1}
d_{ij}\frac{t^{j+1} - t_1^{j+1}}{(j+1)(t-t_1)}.
\end{equation}
The subtraction point $Q_1^2$ will be chosen below to 
ensure that higher dimension condensate contributions 
are safely negligible at scales $Q^2\ge Q_1^2$.

\subsection{FESR analysis}\label{sec:FESR}

One of the drawbacks of the previous lattice analysis, 
Refs.~\cite{Shintani:2008ga,Shintani:2010ph}, involving one
of the current authors, was the low scale used, $1.2\ {\rm GeV}<Q<1.8\ {\rm GeV}$. The lattice data being analyzed showed clear 
signs of $D\ge 4$ contributions, which were handled by
fitting the condensate $C_4$ (which, in addition to the quark
condensate, taken from other lattice calculations, involved the
effective gluon condensate, which was not known from 
other sources) and fixing the analysis window such 
that neglecting $D=6$ (and higher) contributions was
self-consistent.

\enlargethispage{-20mm}

As noted above, there is a potential systematic problem with 
this analysis strategy, since it relies on the implicit
assumption that, as one lowers $Q^2$ from high values,
contributions from the non-perturbative
condensates will turn on one at a time, so that a window of
$Q^2$ will exist in which contributions from $D>4$
condensates will be negligible relative to that of the 
lowest dimension ($D=4$) condensate, with $D=6$ and higher
contributions becoming numerically relevant, one at a time 
as $Q^2$ is lowered, only at yet lower scales. Unfortunately,
FESR results for the higher $D$ condensates do not 
support this picture{\footnote{It is an advantage of the 
FESR framework that, through the freedom of weight
choice, the relative sizes of different higher-dimension OPE
contributions can be easily varied, making it possible to 
fit the higher $D$ condensates, $C_D$, to data. This 
freedom is not available in the approach relying only on
lattice data for the HVP at Euclidean $Q^2$. Thus, while the
lattice approach has many advantages over the FESR one, we
will still need FESR input to help us determine at what $Q^2$ 
potential systematic problems associated with higher $D$ 
contributions become safely negligible in the lattice
approach.}}. Instead, one finds $C_{D\ge 4}$ which generate
higher $D$ contributions to $D(Q^2)$ showing no sign of
convergence with increasing dimension at the scales used in Refs.~\cite{Shintani:2008ga,Shintani:2010ph}. 
This is true whether one considers (i) the results for 
the $C_D$ obtained in Ref.~\cite{Maltman:2008nf} from 
analyzing the vector channel data neglecting DVs, 
(ii) the analogous $V+A$ channel $C_D$ obtained in
Ref.~\cite{Boito:2014sta} in a framework including a model
for DVs, or (iii) the $C_D$ obtained by performing an 
extension, analogous to that carried out for the $V+A$
channel, of the vector channel analysis (including DVs)
reported in Ref.~\cite{Boito:2014sta} to generate
additional $C_D$ beyond the $C_6$ and $C_8$ results reported 
in that paper. These results, moreover, all show strong 
cancellations amongst terms of different dimension having 
different signs, leading to sums of non-perturbative
contributions with much weaker $Q^2$ dependence
than that of the individual terms in the sum.

\begin{figure}
\includegraphics[scale=0.45]{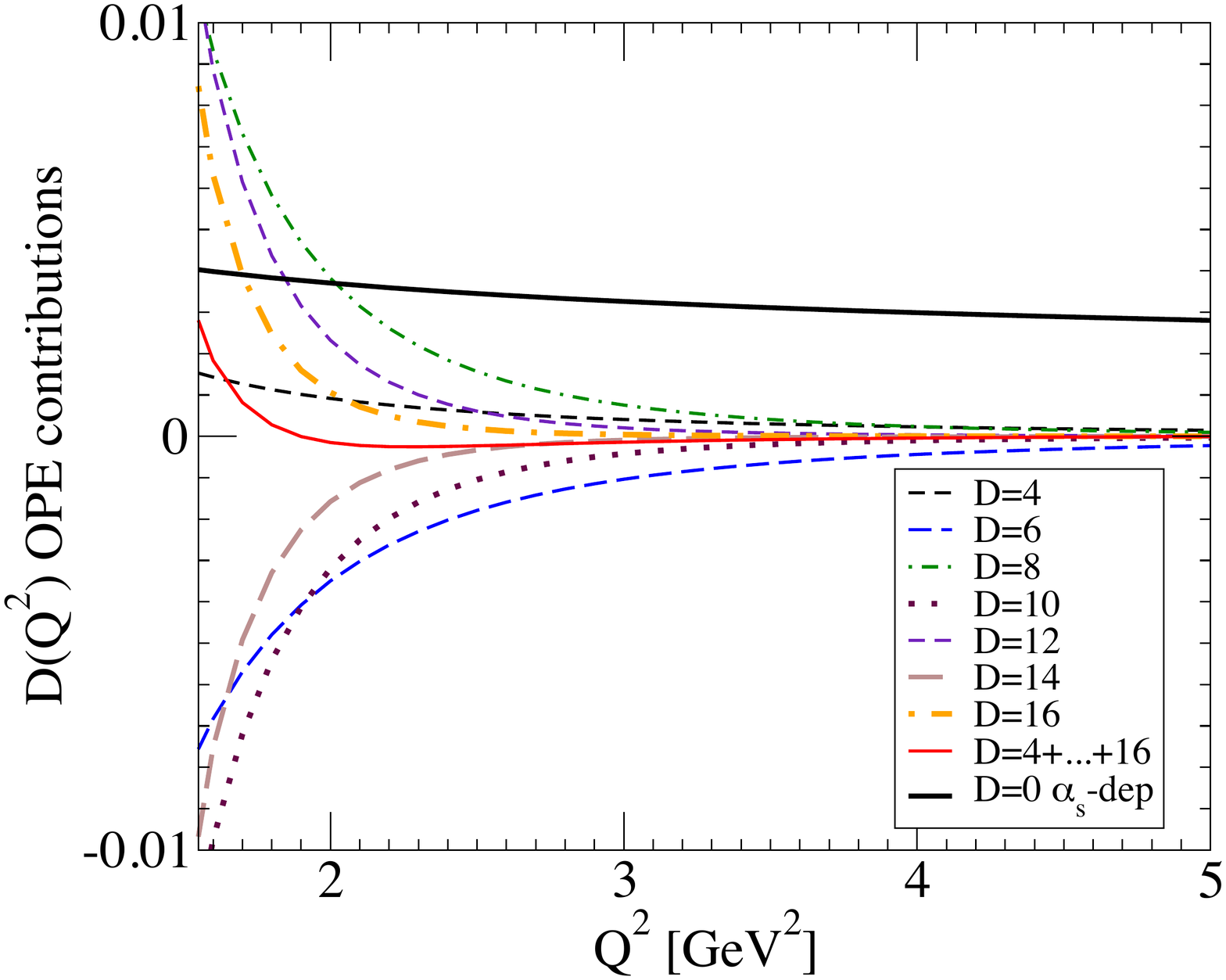}
\caption{$ud$ V channel $\alpha_s$-dependent $D=0$
and non-perturbative $D=4,\cdots ,16$ OPE contributions to
$D(Q^2)$, using $\alpha_s$ and central $C_{D>4}$ value input
described in the text. The $D=0$ curve shows the sum of
the $\alpha_s$-dependent $D=0$ contributions only, with the
$\alpha_s$-independent parton-model contribution removed.}
\label{higherDadler}
\end{figure}
These points are illustrated in Figure~\ref{higherDadler}.
The figure shows the $D=4$ through $16$ contributions 
to $D(Q^2)$, obtained using the central $C_D$ values of 
case (iii) above, together with their much more weakly
$Q^2$-dependent sum. The $Q^2$ range displayed extends down
to $1.55\ {\rm GeV}^2$, the lower edge of the fit window of
Refs.~\cite{Shintani:2008ga,Shintani:2010ph}. Also shown, for 
comparison, is the $\alpha_s$-dependent part of the 
5-loop-truncated perturbative ($D=0$) contribution,
evaluated, for definiteness, using the fixed-scale
expansion form with 3-flavor $\alpha_s$ corresponding to the 
central 5-flavor PDG result, 
$\alpha_s (M_Z)=0.1181$. It is evident that
no single condensate dominates the $D=4$ through $D=16$ sum 
for any of the $Q^2$ shown, and that there is no evidence 
of any convergence with dimension in the low-$Q^2$ region. 
(The upturn of the $D=4$ through $D=16$ sum with $Q^2$ 
at the lowest $Q^2$ shown is thus an artifact of
the truncation of the series, and has no physical meaning.)

Given this evidence for the lack of convergence with $D$ of 
the $D\ge 4$ series, and the failure of any single 
low-dimension condensate to dominate the sum, the only 
safe approach to using lattice HVP data to fix $\alpha_s$ 
is to work at scales $Q^2$ sufficiently large that all
$D\ge 4$ contributions are simultaneously small relative 
to the $\alpha_s$-dependent $D=0$ contributions of 
interest for determining $\alpha_s$. Working at lower 
$Q^2$, where FESR results predict dangerous higher-$D$
effects are almost certainly present, and almost
certainly not amenable to being fitted and removed, makes 
unavoidable a potentially significant theoretical 
systematic uncertainty impossible to successfully quantify
using lattice data alone. We deal with this issue by using 
FESR results to identify $Q^2$ sufficiently large that
all of the estimated individual $D\ge 4$ non-perturbative
contributions to $D(Q^2)$ are safely small, and restrict 
our analysis to $Q_1^2$ and $Q^2$ lying in this region.
Scales above $\sim 3$ to $3.5\ {\rm GeV}^2$ appear 
sufficient for this purpose.

\section{The lattice HVP}\label{sec:lattice}

\begin{table}[h!]
\centering
\begin{tabular}{ c   ccc | ccc | c  }
\toprule
& \multicolumn{3}{|c|}{coarse} & \multicolumn{3}{|c|}{fine} & superfine \\
\hline
extent & \multicolumn{3}{|c|}{$24^3\times64\times32$} & \multicolumn{3}{|c|}
{$32^3\times64\times32$} & $32^3\times64\times32$ \\
$a^{-1}$ (GeV) & \multicolumn{3}{|c|}{1.7848(50)} & \multicolumn{3}{|c|}
{2.3833(86)} & 3.148(17) \\
$am_l$ & \multicolumn{1}{|c}{0.005} & 0.01 & 0.02 & 0.004 & 0.006 & 0.008 &
0.0047 \\
$m_\pi$ (GeV) & \multicolumn{1}{|c}{0.33} & 0.42 & 0.54 & 0.28 & 0.33 & 0.38 &
0.37 \\
measurements & \multicolumn{1}{|c}{685} & 110 & 109 & 510 & 352 & 80 & 920 \\
$am_{\rm res}$ & \multicolumn{3}{|c|}{0.003076(58)} & \multicolumn{3}{|c|}
{0.0006643(82)} & 0.0006296(58)\\
$Z_V$ & \multicolumn{3}{|c|}{0.71408(58)} & \multicolumn{3}{c|}
{0.74404(181)} & 0.77700(8) \\
\botrule
\end{tabular}
\caption{Simulation parameters for the ensembles used in this 
study. $a^{-1}$ and $Z_V$ have been measured in \cite{Blum:2014tka}.}
\label{tab:simparams}
\end{table}

In this work we will use $n_f=2+1$ dynamical Domain Wall 
fermion (DWF) configurations generated by the RBC/UKQCD 
collaborations and extensively detailed in 
Refs.~\cite{Arthur:2012yc,Blum:2014tka}. The relevant ensemble
parameters are listed in Tab.~\ref{tab:simparams}. Since we 
work at the high scales suggested by the FESR analysis of the previous section, where the D=2 series and higher terms in the OPE will be negligible, we expect to be able to safely 
average the results for $\alpha_s$ obtained from ensembles
with different masses. We have checked that these results
are, indeed, consistent within one-sigma errors at the momentum scales used here.

\subsection{Lattice computation}

We start by considering the lattice analog of the continuum 
HVP of Eq.~\ref{eq:pidef} with flavour $ud$ content.

One can define a conserved vector current in the DWF
discretisation~\cite{Kaplan:1992bt,Furman:1994ky} with finite 
fifth dimension extension $L_s$, via
\begin{eqnarray}\label{background:eq:ccvector}
\mathcal{V}_\mu(x) &=& \sum_{s=1}^{L_s} \frac{1}{2}
\bigg( \bar\psi(x+a\hat\mu,s)\left(1+\gamma_\mu\right)
U_\mu^\dagger\left(x+a\frac{\hat{\mu}}{2}\right)\psi(x,s) \nonumber\\
&&\qquad -  \bar\psi(x,s)\left(1-\gamma_\mu\right)U_\mu
\left(x+a\frac{\hat{\mu}}{2}\right)\psi(x+a\hat\mu,s) \bigg) ,
\end{eqnarray}
where $U$ is an $SU(3)$ link matrix. We consider the 
two-point function involving the product of this conserved 
current and the local vector current 
$V_\nu(x) = Z_V\psi(x)\gamma_\nu \psi(x)$. This
combination has been shown to have no contact
term~\cite{Shintani:2008ga,Shintani:2010ph,Boyle:2011hu}.

We find it natural to use the definition of lattice momentum 
that satisfies the lattice Ward identity~\cite{Karsten:1980wd},
\begin{equation}
\sum_\mu a\hat{Q}_\mu e^{iaQ_\mu/2} \Pi_{\mu\nu}(Q) = 0,
\quad a\hat{Q}_\mu = 2\sin(aQ_\mu/2),
\end{equation}
where $Q_\mu = \frac{2\pi n_\mu}{L_\mu}$, with 
$n_\mu = [-\frac{L_\mu}{2a},
\dots ,0 ,\dots ,\frac{L_\mu}{2a}-1 ]$ denoting the Fourier 
modes, and $L_\mu$ is the lattice extent in the $\mu$ 
direction.

\subsection{Projection and cylinder cut}

The Lorentz structure of the lattice two-point function 
should be equivalent to that of the continuum version, 
Eq.~(\ref{eq:lorentz}), up to lattice artifacts from the 
action and from the breaking of rotational symmetry
$O(4)$ under the discrete hyper-cubic group $H_4$. Including 
such artifacts, the lattice two-point function is expected to 
take the form \cite{Shintani:2008ga,Shintani:2010ph},
\begin{equation}\label{eq:higher_order}
\Pi_{\mu\nu}(\hat{Q}:a^2) = \Pi_{\mu\nu}(\hat{Q}) + \sum_{n,m:n+m\geq 2}
C_{nm} a^{n+m}\hat{Q}_\mu^{n}\hat{Q}_\nu^{m} + \sum_{n=1} C_n \delta_{\mu\nu} (a^2\hat{Q}^2)^n
\end{equation}
To limit the impact of lattice artifacts, we utilize momentum 
reflections when projecting out the transverse piece, $\Pi$:
\begin{equation}\label{background:eq:pilatproj}
\Pi(\hat{Q}^2) = \frac{1}{12}\sum_{\mu}\sum_{\nu\neq\mu}
\frac{ \Pi_{\mu\nu}( \hat{Q} ) - 
\Pi_{\mu\nu}( r_\mu \hat{Q}) }
{ 2\hat{Q}_\mu \hat{Q}_\nu },
\end{equation}
where $r_\mu$ is a reflection operator for $Q_\mu$ in the 
$\mu$ direction, $r_\mu Q_\mu = -Q_\mu$. By design, this 
removes all $(n+m)$-even combinations of hypercubic 
artifact terms from $\Pi$.

In addition, to further minimize $O(4)$-breaking effects, we 
perform the reflection projection on purely off-axis,
non-zero momenta that have been filtered by a cylinder 
cut procedure~\cite{Leinweber:1998im}. We accept only 
momenta lying within a radius ($aw$) of a body-diagonal
vector, $n_\mu$, with norm $1$,
$ n_\mu=(\pm 1,\pm 1,\pm 1,\pm 1)/2$, explicitly:
~\cite{PhysRevD.63.094504},
\begin{equation}\label{background:eq:clynder}
\vert aQ_\mu - ( aQ\cdot n ) n_\mu \vert < aw\frac{2\pi}{L}.
\end{equation}
By performing the cylinder cut before we convert our momenta
to physical units we obtain a consistent physical radius. We 
find that a very narrow cylinder width, $aw=0.24$, gives 
smooth results with very limited hyper-cubic artifacts while
still providing plenty of data points to use in performing
fits, as can be seen in Fig.~\ref{fig:data}.

\subsection{Modelling cut-off effects}

We will be attempting to fit the $n_f=3$ series for $\Delta$
(Eq.~\ref{eq:delta_series}) to our lattice data. In principle this fit would involve only one parameter, $\alpha_s(\mu )$, 
if our lattice ensembles were very close to the continuum 
limit, $a^2\rightarrow 0$. With the ensembles available,
dictated by current lattice technology, we must, however, 
also fit residual lattice-spacing artifacts.

Considering the fact that we have projected out several 
combinations of the possible discretisation effects, and have limited the magnitude of the surviving $O(4)$-breaking 
terms, we postulate that the momentum-dependent lattice
corrections will be the rotation-preserving terms 
$C_1\;a^2\hat{Q}^2$ and $C_2\;(a^2\hat{Q}^2)^2$. Such 
effects, if present in our data, will be evident through 
linear and quadratic dependences on $\hat{Q}^2$. In
principle, even higher-order corrections could be present.
The data quality, however, does not permit us to
identify and fit an additional $(a^2\hat{Q}^2)^3$ term.

Another possible cut-off effect in our data, which we will 
fit and remove, is a correction to the continuum coupling
itself,
\begin{equation}
\frac{\alpha_s(\mu:a^2)}{\pi} = \frac{\alpha_s(\mu)}{\pi}
\left( 1 + C_{\alpha} a^2\right).
\end{equation}

Collecting all these correction terms, our fit to $\Delta$ is 
performed using the fit form
\begin{equation}\label{eq:full_fit}
\Delta\left( Q_1^2,Q^2,\mu^2,\frac{\alpha_s(\mu:a^2)}{\pi}\right) +
C_1 \frac{\left( a^2\hat{Q}_1^2 - a^2\hat{Q}^2 \right) }{t_1-t}
+\, C_2 \frac{\left( (a^2 \hat{Q}_1^2)^2 - (a^2 \hat{Q}^2)^2 \right) }{t_1-t},
\end{equation}
where $\Delta\left(Q_1^2,Q_2^2,\mu^2,\frac{\alpha_s(\mu:a^2)}{\pi}\right)$ is the continuum perturbative expression for 
$\Delta$ (Eq.~\ref{eq:delta_series}), with lattice correction to the coupling. We will
choose several subtraction points, $\hat{Q}_1^2$, and fit our 
data over a range of $\hat{Q}^2>\hat{Q}_1^2$. We fit both 
the coupling and all of the correction terms simultaneously.
Our final fit thus has only four-parameters:
$\alpha_s(\mu),\, C_1,\, C_2$ and $C_\alpha$.

\section{Results}\label{sec:results}

\begin{figure}[ht!]
\includegraphics[scale=0.45]{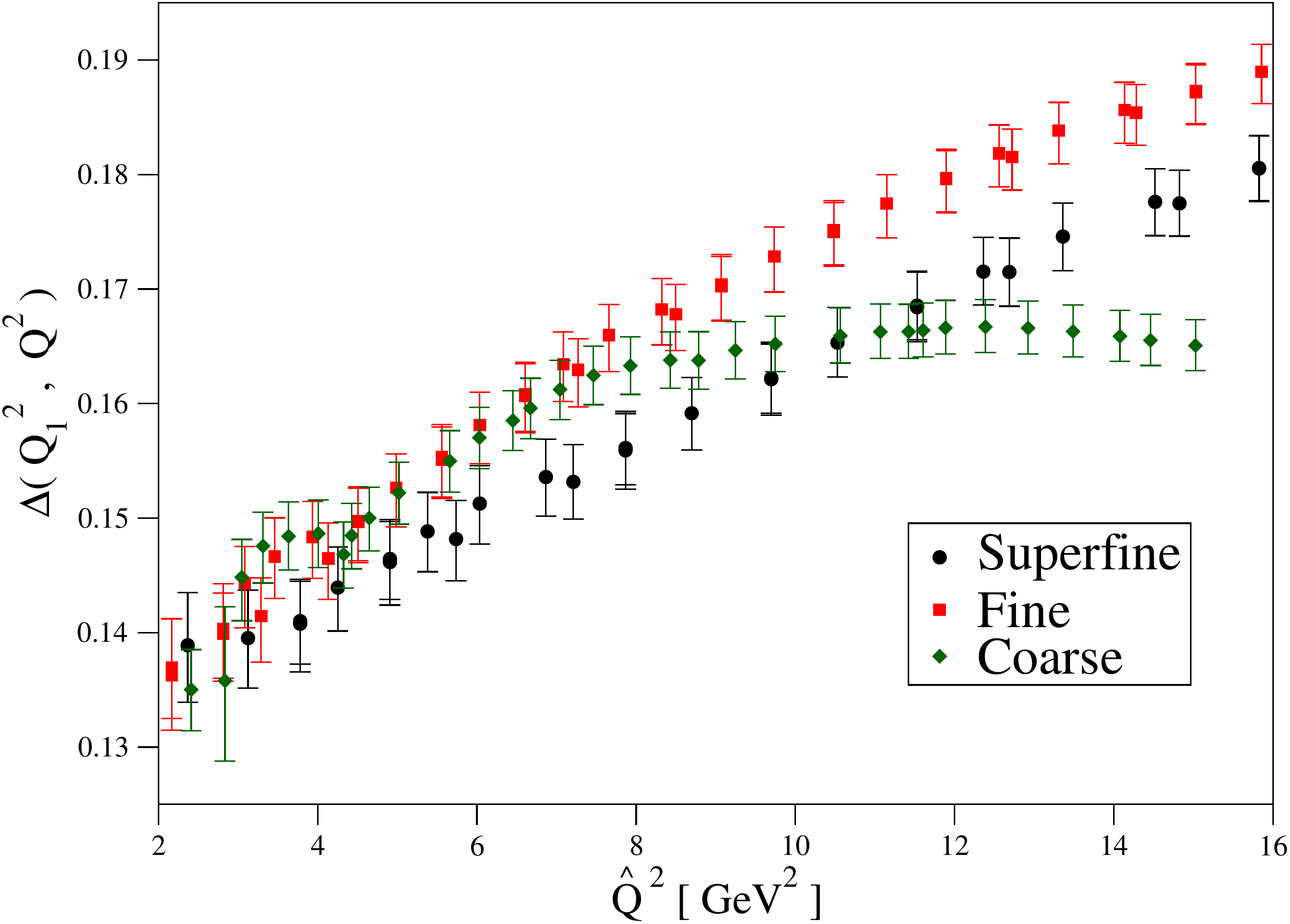}
\caption{Plot of our results for the quantity $\Delta$ against 
$\hat{Q}^2$ with $\hat{Q}_1^2$ values from Tab.~\ref{tab:single_fit}. 
}\label{fig:data}
\end{figure}

We first consider fitting our data using a single renormalisation 
scale $\mu$. We will find that the 
single-scale approach produces a large, unphysical, dependence 
of $\alpha_s$ on the scale $\mu$, and fix this problem using an
alternate multiple-scale approach. Throughout this section all 
errors come from the bootstrap.

Fig.~\ref{fig:data} illustrates the quality of our data for
the quantity $\Delta$ once the reflection projection and
cylinder cut procedures have been applied. It is evident 
that a large linearly-rising behaviour in $\hat{Q}^2$ remains
for all ensembles and that higher-order effects become 
significant for the coarse ensemble, and, to a lesser extent,
for the fine ensemble data, at large $\hat Q^2$.

Throughout this work we will perform what is commonly called an 
``uncorrelated'' or weighted fit to our data since we could not find 
suitable fit ranges where correlated fits were possible. This 
means that, while remaining a useful metric for goodness of fit,
the nominal $\chi^2/dof$ should not be taken as having the usual
statistical interpretation it would have for a correlated fit.
A nominal $\chi^2/dof$ of order 1 or greater will, however,
indicate a poor fit to our data.

\subsection{Single-scale analysis}

We choose to work with the representative fit-ranges shown 
in Tab.~\ref{tab:single_fit}. For now, we are forced to use 
somewhat lower than ideal values of $\hat Q_1^2$ to accommodate 
the coarse ensemble. This is to ensure the subtraction point $\hat Q_1^2$ 
is small enough that cut-off effects will be small at $\hat Q_1^2$,
allowing us to fit the cut-off effects associated with the higher
scale $\hat Q^2$ using their $\hat Q^2$ dependence. We also aim to choose 
subtraction points $\hat Q_1^2$ which are similar for all the ensembles,
subject to the constraint that the $\hat Q_1^2$ for each corresponds
to a Fourier mode.

\begin{table}
\begin{tabular}{ c | c | c }
\toprule
ensemble & $\hat{Q}_1^2$ $[\text{GeV}^2]$ & $\hat{Q}^2$ range 
$[\text{GeV}^2]$\\
\hline
superfine & 2.65 & $3.78 \rightarrow 14.83$ \\
fine & 2.44 & $3.94 \rightarrow 9.74$ \\
coarse & 2.68 & $3.63 \rightarrow 5.02$ \\
\botrule
\end{tabular}
\caption{Fit range parameters for the single-scale and multiple-scale investigations. The 
specific values relate to particular Fourier modes in the data 
accessible after the cylinder cut.}\label{tab:single_fit}
\end{table}

\begin{figure}[ht!]
\centering
\includegraphics[scale=0.35]{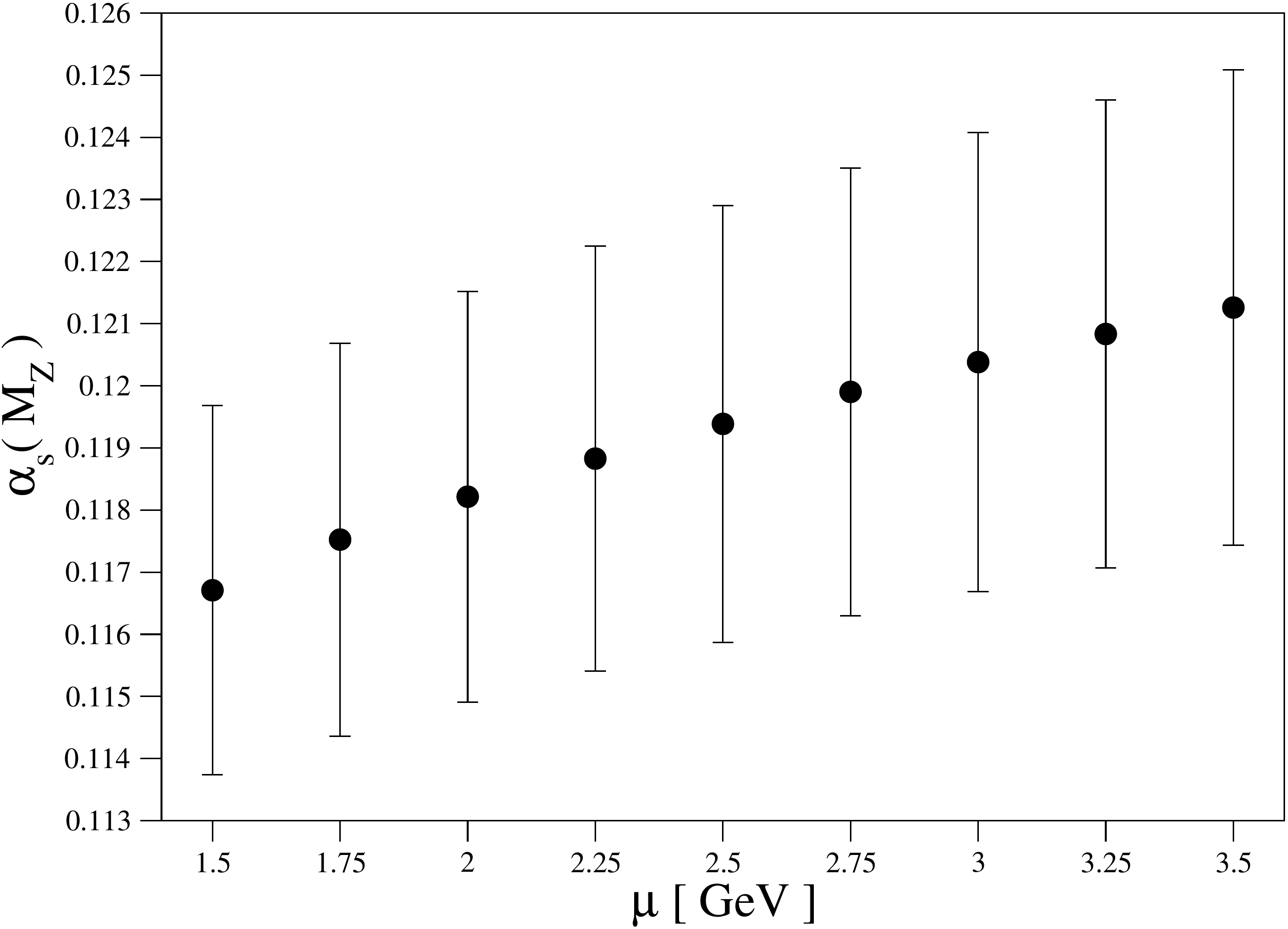}
\caption{$\mu$-dependence of $\alpha_s(M_Z)$ from the single-scale approach.}\label{fig:single_scale}
\end{figure}

In Fig.~\ref{fig:single_scale} we show the $\mu$-dependence 
of our single-scale analysis results for  $\alpha_s(M_Z)$. Since the fit range is fixed, and
the residual, truncation-induced $\mu$-dependence is small in the continuum limit,
the results should be essentially independent of $\mu$ if the single-scale analysis
has been successful in bringing the artifacts under control.

\enlargethispage{3mm}

What we see instead from Fig.~\ref{fig:single_scale} is that the 
single-scale approach, when applied to our data, produces results with
a large-$\mu$ dependence. The $\mu$-dependence is, moreover,
much greater than the residual $\mu$-dependence 
that exists in the continuum truncated perturbative expression\footnote{The single-$\mu$ analysis leads to larger values of $\alpha_s(M_Z)$ with increasing $\mu$, whereas the perturbative expression tends to give a slightly decreasing value of $\alpha_s(M_Z)$.}.

All of the fits underlying the results of Fig.~\ref{fig:single_scale} 
have a nominal $\chi^2/dof<1$, so it is not that the fits do not describe 
the data well. Rather the fits lack sufficient information to successfully
differentiate between the effects of the continuum coupling and
the artifact terms. Similar unphysical $\mu$-dependence is also 
observed for one or more of the artifact coefficients.

One might think that the results of Fig.~\ref{fig:single_scale} are 
more or less consistent within error, but this is not true, given the
very strong correlation among the data points.

\subsection{The multiple-scale analysis}

The strong $\mu$ dependence of the single-scale analysis seen in 
Fig.~\ref{fig:single_scale} is clearly unphysical. A good quality 
fit in a single-scale analysis of our data is thus insufficient to allow both
the continuum coupling $\alpha_s(\mu)$ and the artifact parameters to be reliably constrained. The fact that the unphysical
$\mu$ dependence is large, however, indicates that it must be
possible to further constrain the analysis using the known smallness
of the residual scale dependence of the five-loop-truncated continuum
$D=0$ series.

We thus extend the analysis to perform fits using multiple 
renormalisation scales $m$. These are run to a common renormalisation scale 
$\mu$ using the continuum expression of Eq.~\ref{eq:alpha_run}. 
We use the values $m=2.0,\, 2.25,\, 2.5$ and $2.75$ GeV for the
alternative scale, and for this analysis will vary the renormalisation scale $\mu$ between $1.5$ and $3.5$ GeV as we did for the single-scale analysis for comparison. We will also use the same fit ranges and subtraction points as the single-scale analysis above.

\begin{figure}[h!]
\centering
\includegraphics[scale=0.35]{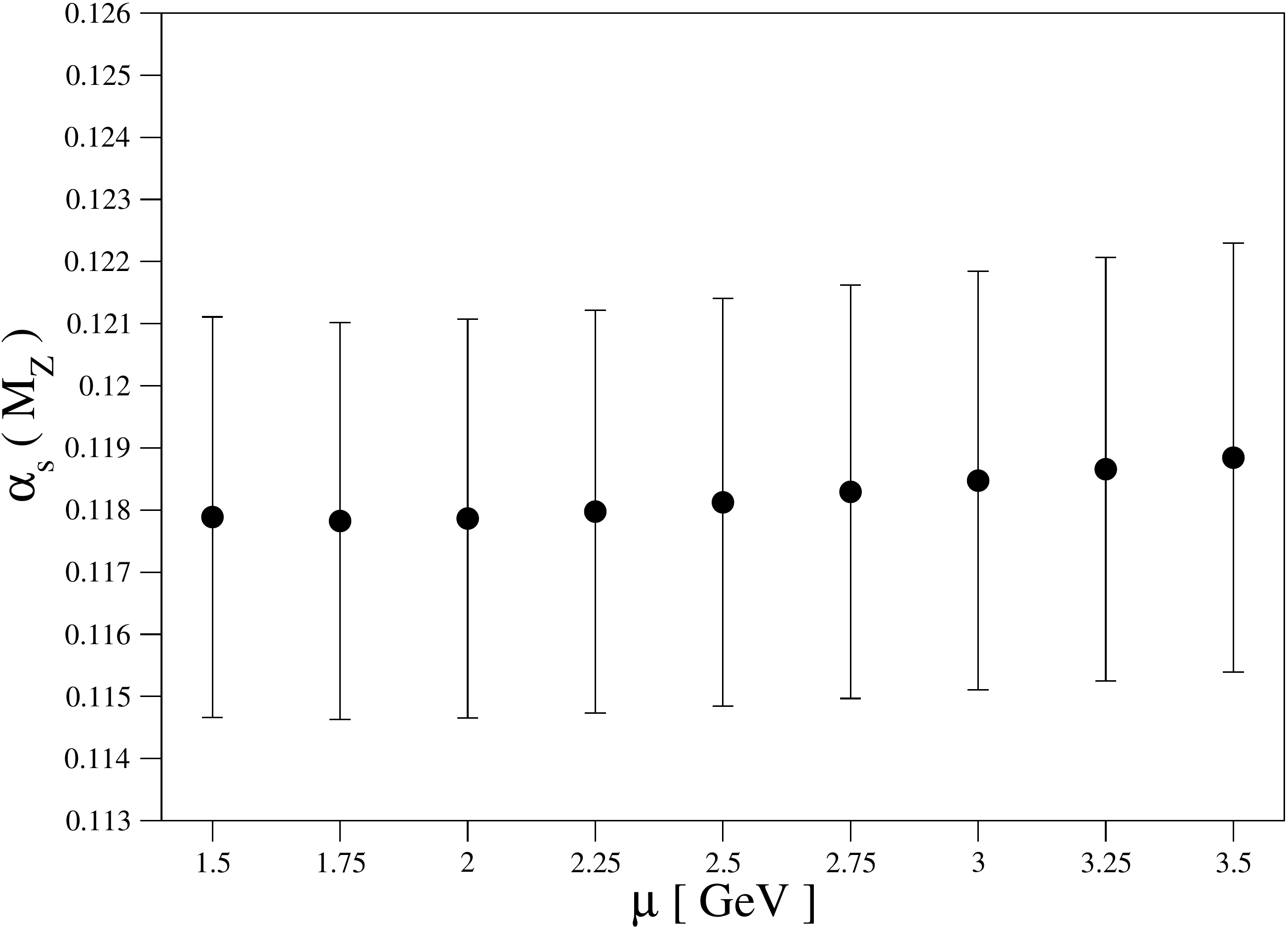}
\caption{$\mu$-dependence of $\alpha_s(M_Z)$ from the multi-scale approach}\label{fig:multi_scale}
\end{figure}

We investigate whether our simultaneous multiple-scale fitting fixes 
the spurious $\mu$-dependence observed in the single-scale analysis
results by again running the coupling to the common five-flavour 
scale $M_Z$. The multiple-scale results for $\alpha_s(M_Z)$ is shown in Fig.~\ref{fig:multi_scale}, for a typical choice of 
fit ranges.

If we compare Figs.~\ref{fig:single_scale} and \ref{fig:multi_scale},
we see that indeed the multiple-scale analysis strongly reduces
the unphysical $\mu$ dependence of the coupling. This is also the case for all of the artifact parameters, indicating that the lattice artifacts
have now been brought under good control over a wide range of scales.

\subsection{Eliminating the coarse data}

\begin{figure}[ht!]
\centering
\includegraphics[scale=0.35]{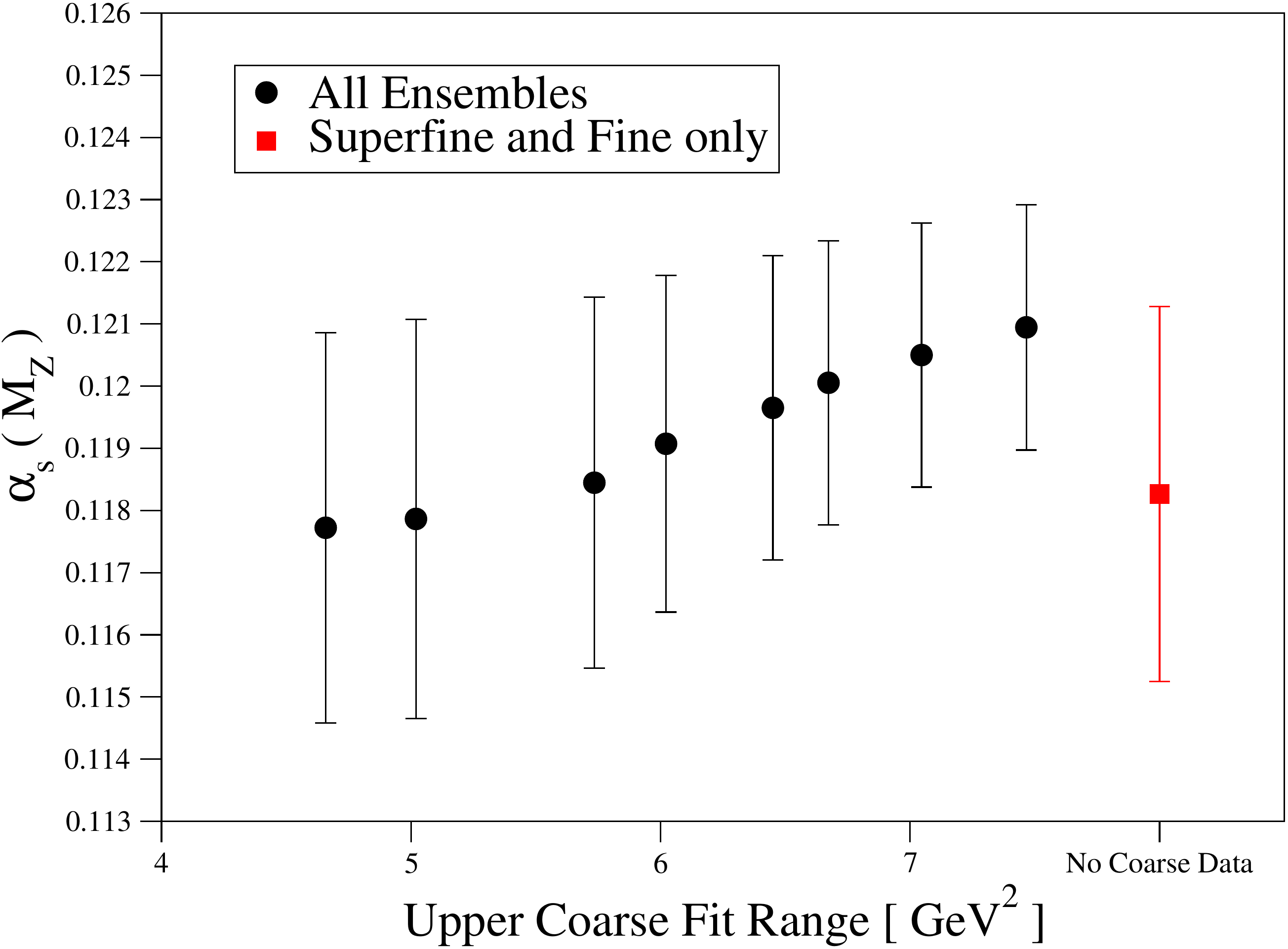}
\caption{Plot of our value of $\alpha_s(M_Z)$ versus the upper bound in 
$\hat{Q}^2$ of the coarse ensemble fit window, with fixed 
lower bound, in comparison to the result obtained with the coarse 
ensemble data entirely removed from the fit.}\label{fig:coarse_elimination}
\end{figure}

As seen in Fig.~\ref{fig:data}, the results from the coarse ensemble have 
the strongest lattice artifacts. Moreover, to safely fit the data only 
a small window in $\hat{Q}^2$ is suitable. If we consider the same set of 
$\hat{Q}_1^2$ as before (Tab.~\ref{tab:single_fit}) and compare 
the results of the multi-scale fit using just the superfine and fine 
ensemble data to those of the corresponding fit using all three ensembles,
we see, by varying the upper bound of the coarse-ensemble's $\hat{Q}^2$ 
fit window, that the coarse ensemble contributes little to the overall 
fit apart from marginally increasing the statistical uncertainty in the region where the result is stable. This 
is illustrated in Fig.~\ref{fig:coarse_elimination}.

As Fig.~\ref{fig:coarse_elimination} shows, in the (small) fit 
window for the coarse ensemble (with the superfine and fine fit 
windows fixed), in the region where the fit appears to be stable 
the result is consistent with the result of the fit using only 
superfine and fine data but has slightly larger errors. We conclude that
cut-off effects for the coarse ensemble are too large to make
employing this ensemble useful for this study. We thus eliminate 
the coarse ensemble from our fits, and base our final results 
on fits involving only the superfine and fine data.

\subsection{Multiple \texorpdfstring{$\hat{Q}_1^2$}{Q1} choices}

\begin{table}[ht!]
\begin{tabular}{ c | c | c | c }
\toprule
ensemble & $\hat{Q}_1^2$ $[\text{GeV}^2]$ & Lower $\hat{Q}^2$ range 
$[\text{GeV}^2]$ & Upper $\hat{Q}^2$ range 
$[\text{GeV}^2]$\\
\hline
superfine & 3.78,4.25,4.9 & $5.5 \rightarrow 6.5 $ & $(2\ {\text{GeV}}^2+\text{fine upper bound}) \rightarrow 16\ \text{GeV}^2$ \\
fine & 3.94,4.51,4.99 & $5.5 \rightarrow 6.5$ & $(2\ {\text{GeV}}^2+\text{lower\ bound}) \rightarrow\ 11\ \text{GeV}^2$ \\
\botrule
\end{tabular}
\caption{Fit ranges and subtraction points $\hat Q_1^2$ investigated in the multiple-$\mu$ analyses. Results for the coupling from these varied fit ranges are given in Fig.~\ref{fig:pole_fitrange}.}\label{tab:multi_fit}
\end{table}

Now that we know we can safely omit coarse ensemble data, 
the much larger cut-off scales of the superfine and fine ensembles
make it possible to choose $\hat{Q}_1^2$ in the range where FESR 
results indicate the neglect of $D\ge 4$ contributions should be very 
safe without incurring overly large discretisation effects. 
For our final results, we perform a multiple-scale analysis
with multiple $\hat{Q}_1^2$ choices to further constrain our fits.

We investigate the fit ranges in Tab.~\ref{tab:multi_fit} to understand our fit-range dependence. We vary the lower edge of 
the fit range in steps of $0.5\text{ GeV}^2$ and the upper edge of the fit range in steps of $1\text{ GeV}^2$. We find good stability when varying the lower
bound over this range, keeping the upper fit bound fixed, for 
the fine and superfine ensembles. In general, it is beneficial to 
have a low fit bound as close to $\hat{Q}_1^2$ as possible as this 
reduces the contributions of the terms proportional to $C_1$ and 
$C_2$, and allows the fit to constrain the coupling better.

\begin{figure}[h!]
\centering
{
	\includegraphics[scale=0.35]{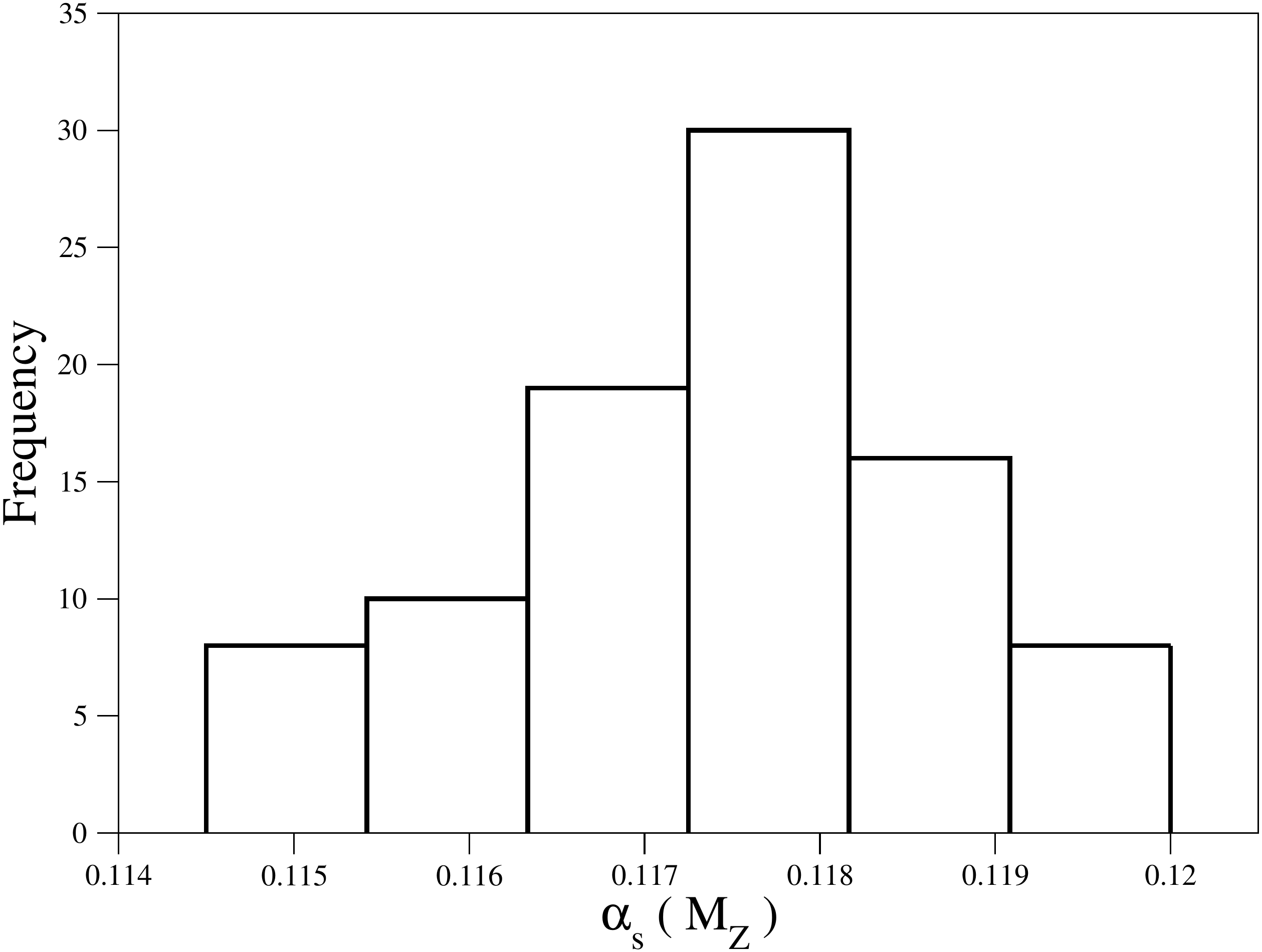}
}
\caption{Histogram of the fit-range-dependence of our final result for 
$\alpha_s(M_Z)$. Results were obtained from a scan over the reasonable fit 
windows in $\hat{Q}^2$ noted in the text.}\label{fig:pole_fitrange}
\end{figure}

A histogram of the central values obtained for $\alpha_s(M_Z)$ from 
a scan over these fit ranges is shown in Fig.~\ref{fig:pole_fitrange}. 
We see that some fit-range-dependence does remain, and this will 
need to be incorporated in our systematic error estimate. Since 
the distribution is skewed, the systematic error will be asymmetric.

Our central result, corresponding to the maximum of the histogram,
is given by the $\hat{Q}^2$ fit bounds of $6.04\text{ GeV}^2 
\rightarrow 9.74 \text{ GeV}^2$ for the fine data and $6.03 \text{ GeV}^2 
\rightarrow 14.83 \text{ GeV}^2$ for the superfine. We use the reference
scale $\mu=2 \text{ GeV}$\footnote{With the multiple-$\mu$ 
fitting strategy there is consistency well within errors between 
the results for $\mu =2 \text{ GeV}$ and those for $\mu=3 \text{ GeV}$.}. 

For these fit ranges, we obtain the following fit parameter results,
\begin{equation}\label{eq:result}
\begin{gathered}
\alpha_s(\mu=2\text{ GeV}) = 0.2961(185),\\
C_\alpha = -3.97(36) \text{ GeV}^2,\quad C_1 = 0.2287(57) ,\quad C_2 = -0.0371(16).
\end{gathered}
\end{equation}
This fit has a nominal $\chi^2/dof=0.7(1)$.

\subsection{Error budget}

We have identified several sources of systematic uncertainty and 
have tabulated their effects in Tab.~\ref{tab:systematics}. A more 
in-depth discussion of how these entries were estimated can be found 
in the following subsections.

\begin{table}[h!]
\begin{tabular}{c | c c}
\toprule
Systematic & \multicolumn{2}{|c}{Estimate} \\
\hline
fit range & +0.5\% & -1.8\%\\
lattice spacing error & +0\% & -0.03\% \\
perturbative truncation & \multicolumn{2}{|c}{negligible} \\
higher order lattice artifacts & +0.4\% & -0.4\% \\
\botrule
\end{tabular}
\caption{Systematic error budget on our final five-flavor
result $\alpha_s(M_Z)$}\label{tab:systematics}
\end{table}

\subsubsection{Fit range}

Even though we have tamed the unphysical $\mu$-dependence of the 
single-scale approach by working at multiple scales, we still see 
some dependence on our fit ranges, as shown in Fig.~\ref{fig:pole_fitrange}. 
All of these results correspond to reasonable fit ranges in $\hat{Q}^2$ and a low nominal $\chi^2/dof$. To 
estimate this systematic we will take 
the one-$\sigma$ deviation of the histogram.

\subsubsection{Lattice spacing error}

We cannot propagate the measured distributions for the lattice spacing  directly, so we estimate the the impact of the error on the lattice spacing on our final result as a systematic. We do this by introducing $a^2$ fit parameters for the fine and superfine ensembles into the final fit with a prior width of the errors given in Tab.~\ref{tab:simparams}. This introduces a negligible downward correction to the central value of $\alpha_s(M_Z)$ of the order of 0.03\%.

\subsubsection{Perturbative truncation}

If we vary the loop order in our analysis, using an estimated value 
of $d_5=400$\footnote{We consider this value to be fairly conservative and in the same ballpark as other guesses \cite{Baikov:2008jh,Beneke:2008ad}.} to extend this to six loops (in this case, still using 
just the combination of 4-loop running and 3-loop threshold matching
), we see that our results at three-loop order 
are consistent with those at six-loop order. Perturbative truncation is thus not a significant 
source of systematic uncertainty.

\begin{figure}[h!]
\centering
\includegraphics[scale=0.35]{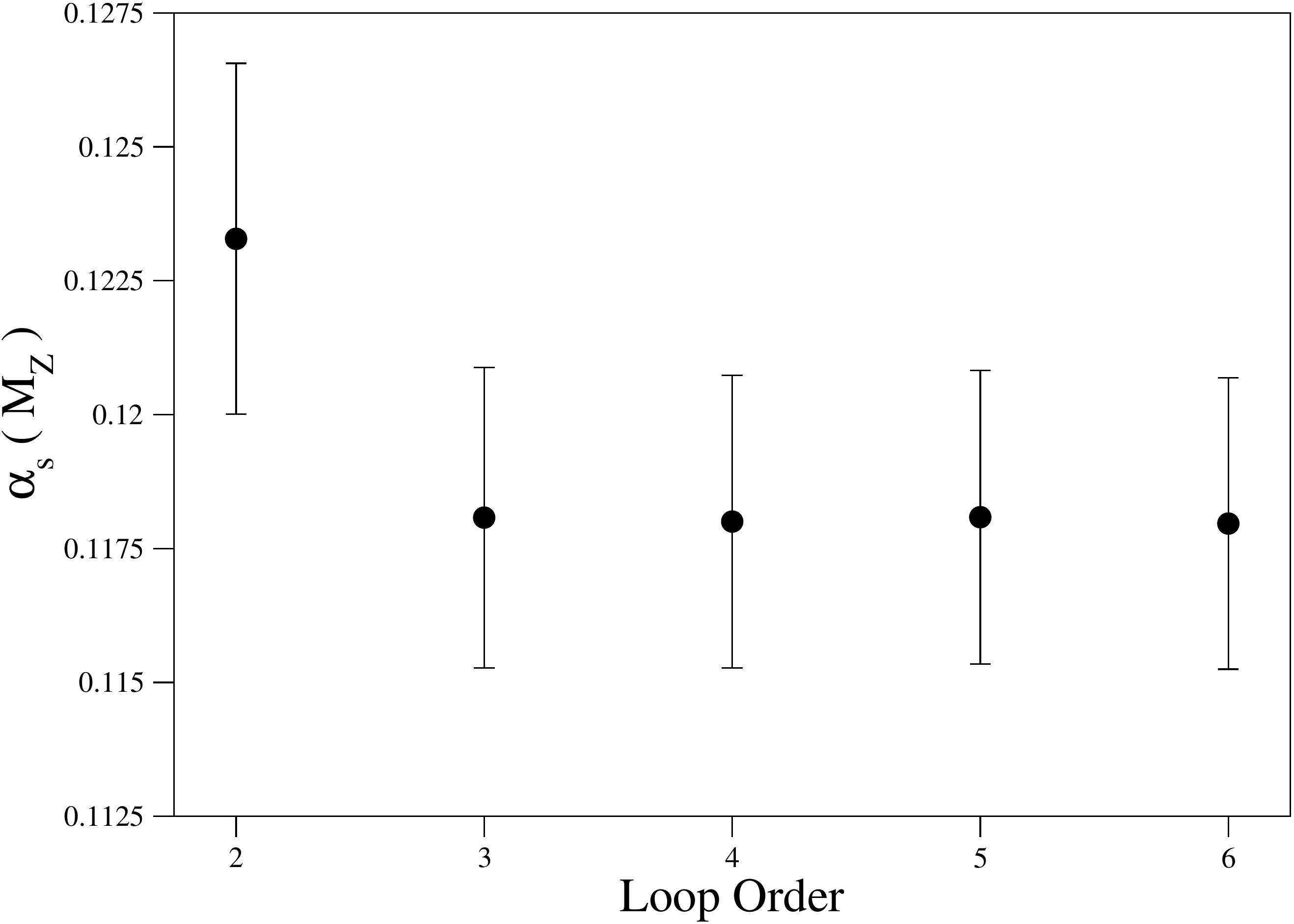}
\caption{Results for $\alpha_s(M_Z)$ with different orders of 
truncation for the $D^{(0)}$ series, running and threshold matching
An estimate of the unknown parameter $d_5$ of 400 was used for the 
six-loop evaluation.}
\end{figure}

\subsubsection{Higher order lattice artifacts}

With our fit parameters (Eq.~\ref{eq:result}), we note that the 
$O(a^2)$ correction term in $\alpha_s(\mu)$ is large even for our superfine 
ensemble. We found we could not stably fit an $O(a^4)$ correction term to our data so we must estimate 
its possible effect. We note that $C_2 \simeq 0.16\,C_1$ and thus expect neglected, higher-order rotation-breaking and rotation-preserving contributions to be smaller than those
included in our fits. This is to be expected from the lack of strong curvature in the fine and superfine data of Fig.\ref{fig:data} within our chosen fit range. 

To estimate the effects of an $a^4$ correction on the coupling $\alpha_s(\mu)$ we put a coefficient of $a^4$ into the fit, with a fixed coefficient of $\pm 1$. Considering the hierarchy represented by the ratio $C_2/C_1$, we believe this to be a reasonably conservative estimate. Studies with more and finer lattice spacings would be required to further test this assumption. The inserted correction term produces small shifts to the central result of $\alpha_s(M_Z)$ of $\pm 0.4\%$.

\subsection{Result at \texorpdfstring{$M_Z$}{Mz}}

Our final result for the five-flavor coupling, $\alpha_s(M_Z)$, is
\begin{equation}
\alpha_s(M_Z) =  0.1181(27)^{+8}_{-22},
\end{equation}
where the first error is statistical and the second contains the systematic 
errors identified in the previous section combined in quadrature.
    
\section{Conclusions}\label{sec:conclusion}

We have revisited the determination of $\alpha_s$ from OPE
fits to lattice data for the flavor $ud$ vector current HVP,
identifying and dealing with several sources of systematic
error not recognized in previous implementations of this 
approach. 

One key problem, of particular relevance to low-scale 
versions of this analysis, is the possibility that multiple
numerically significant higher dimension OPE contributions 
can, through cancellation, produce a sum which mimics 
lower dimension contributions and hence potentially
contaminates the determination of $\alpha_s$. Such effects, 
if present, cannot be fitted and removed using lattice
data at Euclidean $Q^2$ alone. Continuum FESR results
were (i) shown to confirm that this problem was almost
certainly present at the low scales, $Q^2<3\ \text{GeV}^2$, of the previous 
versions of this analysis, then (ii) used further to 
identify those $Q^2$ at which all such higher
dimension OPE contributions should be safely negligible.
For such $Q^2$, the lattice HVP should be dominated by 
dimension $0$ (perturbative) OPE contributions and 
residual lattice artifact contributions, with the latter
to be fitted and removed. We have introduced a momentum-reflection-based
projection method for the lattice HVP and employed a very narrow
cylinder cut to further reduce higher-order rotation-breaking
cut-off effects.

A second key problem is the large, unphysical, $\mu$
dependence observed when employing the (previously used) 
single-scale approach to fitting the HVP data. This problem
was rectified by implementing a multiple-scale analysis strategy.


We find that some residual fit-range dependence does remain, 
even after performing a multiple-scale, multiple-$Q_1^2$ 
analysis. For the lattice spacings of the present study, 
lattice artifacts are not small in the $Q^2$ range employed 
and these had to be carefully modelled to obtain a well-behaved 
result. Additional ensembles finer than the superfine 
ensemble employed here (which will have smaller lattice
artifacts in the same $Q^2$ range) should help to 
further reduce the systematic uncertainty, as well as 
improve statistics. One significant advantage of 
our approach is that simulations at
multiple unphysical (heavy) pion masses can be used
simultaneously and averaged to improve statistics, since, for the $Q^2$ 
in our chosen fit windows, mass-dependent OPE contributions 
are negligible over the range of $m_\pi$ typical of 
modern simulations.

Our final result is consistent with both the PDG 
central value of $0.1181(11)$ \cite{Olive:2016xmw} 
and the FLAG result of $0.1182(12)$ \cite{Aoki:2016frl}. 
Our errors, both statistical and systematic, are 
currently larger, but checking for consistency is important 
for the determination of such a fundamental constant. 
We also believe we have identified the source of the tension 
between the world average and the results of the previous, 
lower-scale version of this analysis \cite{Shintani:2010ph} 
(which gave $\alpha(M_Z)=0.1118^{+16}_{-17}$) and, with our 
improved method, have succeeded in solving the problems
that led to this tension. 
Future improvements to both the
statistical and systematic errors, moreover, appear technically 
straightforward.

We conclude by stressing the close relation of the lattice
HVP approach discussed in this paper to the 
$\tau$-decay-data-based FESR determination of $\alpha_s$.
Both involve an analysis of the same object, the flavor 
$ud$ HVP. The lattice approach, however, has certain
important advantages. In particular, while the somewhat
low scale of the $\tau$ mass makes unavoidable the
presence of higher dimension OPE and DV contributions
in the FESR analysis, DV contributions are exponentially
suppressed at the Euclidean $Q^2$ of the lattice analysis
and, with modern lattices, it is feasible to work at
scales, $Q^2$, where higher dimension OPE contributions
are safely negligible. The price to pay for these
advantages is the presence of residual lattice artifacts
that have to be fit and removed from the lattice HVP
data at these higher scales. This procedure, however,
is subject to systematic improvement through increased
statistics and the use of finer lattices with reduced
artifact contributions. As such, it appears to us
that, in the long run, the lattice approach represents 
the more favourable of the two methods for extracting 
$\alpha_s$ from the light-quark HVP.

\section{Acknowledgements}

This work was supported by the DiRAC Blue Gene Q Shared Petaflop system at the University of Edinburgh, operated by the Edinburgh Parallel Computing Centre on behalf of the STFC DiRAC HPC Facility (www.dirac.ac.uk). This equipment was funded by BIS National E-infrastructure capital grant ST/K000411/1, STFC capital grant ST/H008845/1, and STFC DiRAC Operations grants ST/K005804/1 and ST/K005790/1. We would thank to the members of RBC/UKQCD collaboration for helpful discussions and comments. RJH, RL and KM are sponsored by the Natural Sciences and Engineering Research Council of Canada (NSERC). ES is grateful to RIKEN Advanced Center for Computing and Communication (ACCC) and the Mainz Institute for Theoretical Physics (MITP).

\section{Appendices}

Here we put the relevant series coefficients for the perturbative series we have used in this work. All coefficients are for the $n_f=3$ variants.

\subsection{\texorpdfstring{$\beta$}{Beta}-function coefficients}\label{app:betacoeffs}

\begin{equation}\label{eq:nfthree_beta}
\begin{aligned}
\beta_0 = 9/4 , \quad \beta_1 = 4 ,\quad\beta_2 = 3863/384 , \quad \beta_3 = 47.228040 .\\
\end{aligned}
\end{equation}

\subsection{Running coupling expansion coefficients}\label{app:aseries}

\begin{equation}\label{eq:ascale}
\begin{gathered}
A_{21} = -\beta_0,\\
A_{32} = \beta_0^2,\quad A_{31} = -\beta_1,\\
A_{43} = -\beta_0^3,\quad A_{42} = \frac{5}{2}\beta_0\beta_1,\quad A_{41} = -\beta_2, \\
A_{54} = \beta_0^4, \quad A_{53} = -\frac{13}{3}\beta_0^2\beta_1,
\quad A_{52} = \frac{3\beta_1^2 + 6\beta_0\beta_2}{2},\quad A_{51} = -\beta_3.
\end{gathered}
\end{equation}

\subsection{\texorpdfstring{$D^{(0)}$}{D-zero} series coefficients}\label{app:dseries}

\begin{equation}\label{eq:d0_series}
\begin{gathered}
d_{00} = 1,\quad d_{10} = 1,\\
d_{20} = 1.639821,\quad d_{21} = -2.25,\\
d_{30} = 6.371014,\quad d_{31} = -11.379195, \quad d_{32} = 5.0625,\\
d_{40} = 49.075700,\quad d_{41} = -66.182813,\\\quad d_{42} = 47.404784, \quad d_{43} = -11.390625,\\
d_{50} = d_5,\quad d_{51} = -598.354375, \quad d_{52} = 388.732597,\\ \quad d_{53} = -162.464353, \quad d_{54} = 25.628906.
\end{gathered}
\end{equation}

\bibliography{alphaHVP}{}
\bibliographystyle{h-elsevier}

\end{document}